\documentclass[doublecol,figures]{epl2}
\usepackage{amsmath,amsbsy,amssymb,xspace,enumerate,color,MnSymbol}
\usepackage[plain]{fancyref}
\usepackage[utf8]{inputenc}

\newcommand{\eg}{\textit{e.\,g.},\ }

\newcommand{\THBM}{T_\mathrm{HBM}}
\newcommand{\zetaHBM}{\zeta_\mathrm{HBM}}
\newcommand{\DHBM}{D_\mathrm{HBM}}
\renewcommand{\vec}[1]{\mathbf{#1}}
\renewcommand{\revision}[1]{#1}
\newcommand{\tensor}[1]{\mathsf{#1}}

\newcommand{\SUPMAT}{supplementary online materials\cite{supmat}}

\definecolor{mypurple}{RGB}{153,61,113}
\definecolor{myblue}{RGB}{63,61,153}
\definecolor{myokker}{RGB}{153,140,61}
\definecolor{mygreen}{RGB}{61,153,86}
\definecolor{mymarine}{RGB}{61,90,153}
\definecolor{mycyan}{RGB}{0,255,255}

\title{Generalised Einstein Relation for Hot Brownian Motion}
\author{D. Chakraborty\inst{1,2}
  \and M. V. Gnann\inst{3} \and
  D. Rings\inst{1} \and J. Glaser\inst{1} \and F. Otto\inst{3} \and F. Cichos\inst{2} \and
  K. Kroy\inst{1}\thanks{E-mail:\email{kroy@uni-leipzig.de}} } \shortauthor{D. Chakraborty \etal}

\institute{
  \inst{1} Institute for Theoretical Physics - University of Leipzig, Postfach 100920, 04009 Leipzig, Germany\\
  \inst{2} Institute for Experimental Physics - University of Leipzig, Linn\'estra\ss e 5, 04103 Leipzig, Germany\\
  \inst{3} Max-Planck-Institute for Mathematics in the Sciences,
  Inselstr. 22, 04103 Leipzig, Germany }

\pacs{05.40.Jc}{Brownian motion}
\pacs{05.70.Ln}{Nonequilibrium and irreversible thermodynamics}
\pacs{47.11.Mn}{Molecular dynamics methods}
\pacs{05.40.-a}{Fluctuation phenomena, random processes, noise, and Brownian motion}

\abstract{The Brownian motion of a hot nanoparticle is described by an
  effective Markov theory based on fluctuating hydrodynamics. Its
  predictions are scrutinized over a wide temperature range using
  large-scale molecular dynamics simulations of a hot nanoparticle in
  a Lennard-Jones fluid.  The particle positions and momenta are found
  to be Boltzmann distributed according to distinct effective
  temperatures $\THBM$ and $T_\mathrm{k}$. For $\THBM$ we derive a
  formally exact theoretical prediction and establish a generalised
  Einstein relation that links it to directly measurable quantities.}

\begin{document}

\maketitle

Hot Brownian motion is the Brownian motion of nanoscopic colloidal
particles that have an elevated temperature compared to their
solvent\cite{Rings:2010}. The phenomenon is ubiquitous in modern
biophysical and nanotechnological work, where one often relies on
nanoparticles exposed to laser light as tracers, anchors, or
self-propelling
nanomachines\cite{Ruijgrok:2011,Jiang:2010,Lasne:2006,vanDijk:2006,octeau-etal:2009,Huang:2010,Gaiduk:2010}. In
a number of innovative applications the heating of the particle is
moreover deliberately exploited for detection, manipulation and
surgery on the nanoscale
\cite{bericiaud-etal:2004,radunz-etal:2009,Urban:2009,Qian:2008,Kyrsting:2011}. In
general, heat and vorticity diffuse much faster than the colloidal
particles, typical diffusivities being on the order of
$10^{-7}\,\mathrm{m}^2\,\mathrm{s}^{-1}$ versus $10^{-11}\ldots
10^{-10}\,\mathrm{m}^2\,\mathrm{s}^{-1}$, respectively. On the basis
of this characteristic Brownian scale separation, one may treat the
solvent temperature and velocity distributions as stationary radial
fields $T(r)$, $\vec u(\vec r)$ around the instantaneous particle
position. To a good approximation, the hot Brownian motion of a single
particle can thus be characterised as a stationary nonequilibrium
process.  It is then natural to seek an effective equilibrium
description in terms of Markovian stochastic equations of motion with
effective friction and temperature parameters
\cite{Rings:2010,Rings:2011}.

Such effective parameters are certainly valuable elements of any
quantitative description or application, but it is also clear that
they will be less universal and more context-sensitive than their
conventional equilibrium counterparts. For example, one has to expect
different effective parameters for different degrees of freedom of the
colloidal particle, such as translational and rotational motion, and
they differ not merely by simple geometric or kinematic factors
\cite{Rings:Rotational:2011}. Also momentum and conformational degrees
of freedom turn out to be governed by distinct effective temperature,
as already independently pointed out by Barrat and
coworkers\cite{Joly:2011}. The best one can hope for is thus a
systematic and quantitative understanding of the origin of the
effective temperatures and transport coefficients pertaining to the
observables most relevant in practice. This hope is justified by the
observation that Brownian motion is a mesoscopic phenomenon and as
such allows some coarse-graining over microscopic degrees of
freedom. In other words, the effective temperatures and transport
coefficients of the colloidal particle should emerge form the ``middle
world''\cite{haw_middle-world:2006} of stochastic
thermodynamics\cite{Seifert:2008} and fluctuating
hydrodynamics\cite{Fox:1970a,Fox:1970b,hauge-martin_lof:73} rather
than directly from a much more intricate microscopic description. As a
consequence, the effective parameters may still be expected to be
reasonably insensitive to many of the usually elusive (and often
accidental) microscopic details, such as the precise functional form
of the atomic interactions.

While this is true in principle, the appropriate mesoscopic approach
is not always entirely obvious and straightforward, \emph{a
  priori}. Therefore, it is valuable to have direct access to a
comprehensive microscopic characterisation of some model system. A
standard way to achieve this is via molecular dynamics (MD)
simulations, which, in contrast to real experiments, provide complete
control over the microscopic conditions.  Following the pioneering
work by Alder and Wainwright, there have been extensive investigations
of the microscopic basis of classical fluid dynamics in general, and
of Brownian motion and its transport properties in particular
\cite{Bocquet:1994,Keblinski:2006,Li:2009,Shin:2010}. In the
following, we report on MD simulations of a hot nanoparticle in a
Lennard-Jones solvent. We find that its Brownian motion is
characterised by a set of four distinct effective temperatures, two
for its rotational and translational configurational dynamics and two
for the corresponding momentum or kinetic (k) degrees of freedom.  In
particular, we demonstrate the following statements that hold for both
rotational and translational Brownian motion (the explicit
demonstration for rotational motion is deferred to a companion paper
\cite{Rings:Rotational:2011}).
\begin{enumerate}
\item The characterisation in terms of effective temperatures for the
  configurational and kinetic degrees of freedom, previously
  investigated for a free particle
  \cite{Rings:2010,Rings:2011,Joly:2011}, still holds in presence of
  potential forces. It is largely insensitive to microscopic details
  (\eg to the solubility of the nanoparticle), but the ``kinetic
  temperatures'' are sensitive to the precise heating mechanism.
\item The ``configurational temperature'' $\THBM$ may be expressed in
  terms of the effective friction coefficient $\zetaHBM$ and diffusion
  coefficient $\DHBM$ of the particle via a generalised Einstein
  relation,
\begin{equation}
  \label{eq:GER}
  k_B\THBM=\DHBM\,\zetaHBM\;.
\end{equation}
This allows $\THBM$ to be inferred from directly measurable
quantities. Over a broad range of particle temperatures, \fref{eq:GER}
is found to be in excellent agreement with a formally exact
theoretical prediction, \fref{eq:T_HBM_Manuel}, derived in the
\SUPMAT.
\end{enumerate}
 
\textbf{Theory} The derivation of the effective temperature $\THBM$
that characterises the position fluctuations of a hot Brownian
particle parallels the contraction of the fluctuating Stokes problem,
as established for equilibrium Brownian motion
\cite{hauge-martin_lof:73}. In our analytical calculations, the
solvent is idealised as incompressible and the temperature and
viscosity distributions are approximated by radial fields centred at
the instantaneous particle position, which is well justified for many
experimental realisations.  The crucial quantity in the calculation is
the dissipation function
\begin{equation}
  \label{eq:diss}
  \phi(\vec r)\equiv 2\eta(r)
  \tensor{\Gamma}(\vec{r}):\tensor{\Gamma}(\vec{r})\;,
\end{equation}
which gives the energy dissipated locally by the solvent flow in terms
of the strain rate tensor
$\tensor{\Gamma}\equiv(\nabla\vec{u}+\nabla\vec{u}^T)/2$ for a given
solvent velocity field $\vec u(\vec r)$. Assuming for the fluctuating
part of the solvent stresses the standard equilibrium correlations,
evaluated at the local temperature $T(\vec r)$, our formally exact
calculation, \revision{as detailed in the \SUPMAT,} gives
\begin{equation}
  \label{eq:T_HBM_Manuel}
  \THBM=\frac{\int_VT(\vec{r})\phi(\vec{r})\,\upd^3r}{\int_V\phi(\vec{r})\,\upd^3r}\;.
\end{equation}
\revision{Analytical evaluation of
\fref{eq:T_HBM_Manuel} for constant viscosity and thermal
conductivity---corresponding to the standard velocity and temperature
profiles $4\vec u(\vec r) = -3[\vec U + \hat{\vec r}(\vec
U\cdot \hat{\vec r})]R/r - [\vec U - 3\hat{\vec r}(\vec U\cdot
\hat{\vec r})]R^3/r^3$ and $T(r)=T_0+\Delta T R/r$,
respectively---yields the exact result,
$\THBM = T_0+5\Delta T/12$. Here, $\Delta T$ denotes the temperature
difference between the ambient temperature $T_0$ and the
solvent temperature at the surface of the particle.
The temperature dependence of the viscosity and
the thermal conductivity gives rise to non-universal terms of higher
order in the temperature increment $\Delta T$}. Their explicit evaluation
requires some additional effort. Namely, one has to actually solve the
generalised Stokes problem
\begin{equation}
  \label{eq:Stokes_eq}
  \nabla p = 2\nabla\cdotp\eta\tensor{\Gamma} = 
  2\tensor{\Gamma}\nabla\eta+2\eta\nabla\cdotp\tensor{\Gamma}\,,\quad\nabla\cdotp\vec{u}=0\;,
\end{equation}
for a radially varying viscosity field $\eta(r)$ to explicitly
determine the velocity field $\vec u(\vec r)$, subject to the usual
no-slip boundary condition $\vec u(R)=\vec U$. \revision{Although real fluids,
such as water, are slightly compressible, the assumption
$\nabla\cdotp\vec{u}=0$ is still valid, even if
$\nabla\cdot(\rho\vec{u})\not =0$, as the density variation is assumed
to move stationarily along with the Brownian particle.} From $\vec u(\vec r)$
one then obtains $\phi(\vec r)$ and thus $\THBM$, and the effective
friction coefficient $\zetaHBM$ of the hot Brownian particle.  To
solve \fref{eq:Stokes_eq}, a differential shell method has been
developed that yields exact numerical
results\cite{Rings:2011}. For most practical applications,
this complication can, however, be sidestepped by using an analytical
procedure that produces quite accurate predictions for $\phi(\vec r)$
and $\zetaHBM$ from
\fref{eq:Stokes_eq}\cite{Rings:2010,Rings:2011}. If the temperature
dependence of the solvent viscosity can be parameterised by the
phenomenological relation $\eta(T)=\eta_\infty
\exp[A/(T-T_\mathrm{VF})]$ (as \eg for water), and the density and
thermal conductivity of the solvent can be assumed to be constant, it
yields
\begin{equation}
  \label{eq:pade_approx_VF}
  \THBM \approx T_0 + \frac{5}{12}\Delta T +
  \frac{\ln(\eta_0/\eta_\infty)}{22(T_0 + \Delta T- T_{\rm VF})} \; 
  \Delta T^2 \;,
\end{equation}
which improves eq.~(13) of Rings \etal\,(2010)~\cite{Rings:2010} and
corresponding estimates in Rings \etal\,(2011)~\cite{Rings:2011}
(numerical errors less than $2\%$ for $\Delta T/T_0< 1$). For a more
detailed discussion, including a corresponding expression for the
Lennard-Jones fluid and an explicit formula for the effective
diffusivity, the reader is referred to the \SUPMAT.

\begin{figure}[t]
  \centering
  \includegraphics[width=0.75\linewidth]{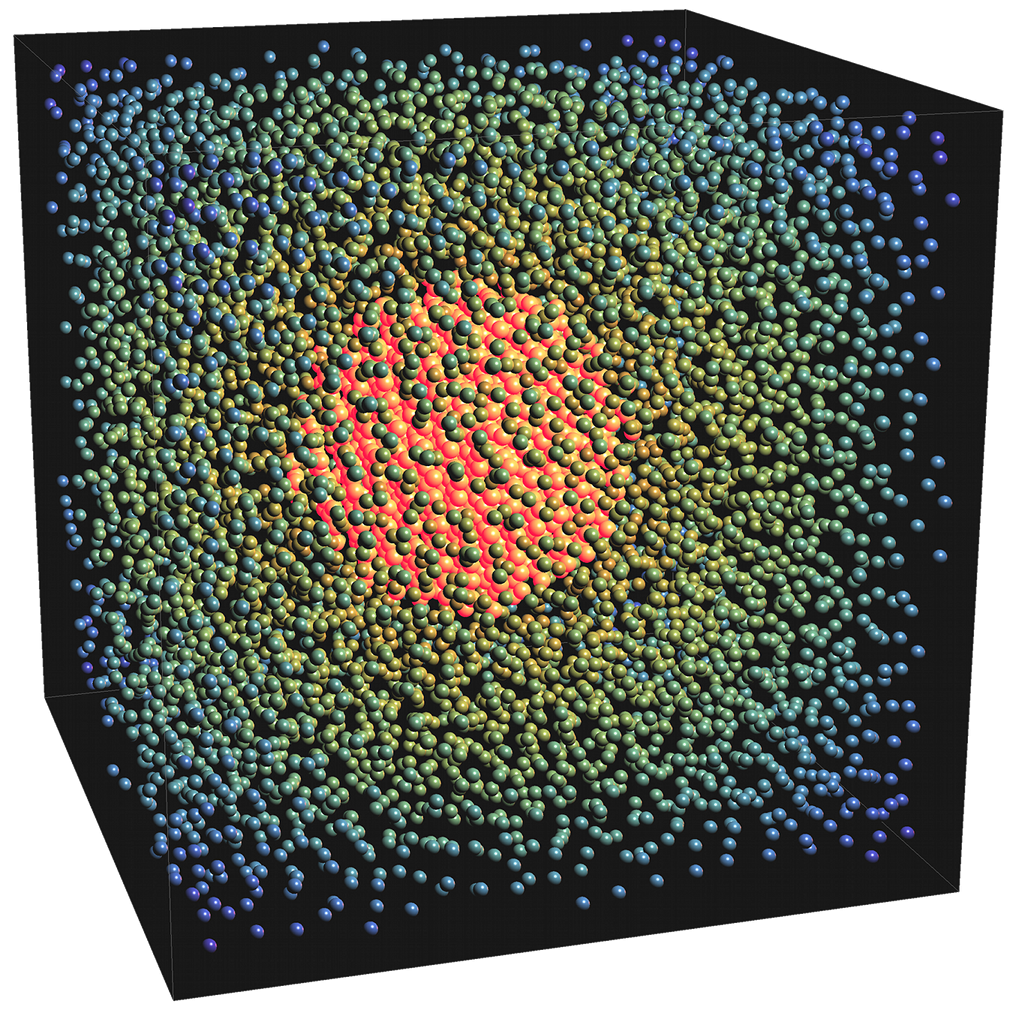}  
  \caption[For LoF]{Snapshot of the simulation: the Lennard-Jones
    atoms in the nanoparticle and the solvent are color-coded in
    order to visualise the temperature gradient.}
  \label{fig:screenshot}
\end{figure}

\textbf{MD simulations} In the simulations, the solvent and the
Brownian nanoparticle are not treated as continuous media but are
themselves made of atoms interacting via the Lennard-Jones radial
pair potential $U(r)=4 \epsilon [(\sigma/r)^{12}-(\sigma/r)^6]$
truncated at $r=2.5 \sigma$. The atoms belonging to the nanoparticle
are additionally bound together by a FENE potential $U(r)=-0.5 \kappa
R_0^2 \log[1-(r/R_0)^2]$ with $\kappa=30 \epsilon/\sigma^2$,
$R_0=1.5\sigma$.  As usual, we measure lengths, times, and energies in
terms of the Lennard-Jones units $\sigma$, $\tau
\equiv\sqrt{m\sigma^2/\epsilon}$ and $\epsilon$, respectively. For
liquid Argon, they correspond to $\sigma=3.405\,$\AA,
$\epsilon/k_\mathrm{B}=119.8\,$K, $m=0.03994\,$kg/mol and $\tau \sim
2\,$ps\cite{Frenkel:2002}. We note that the critical temperature for
the bulk Lennard-Jones fluid with a cutoff of $2.5$ is
$T_c=1.186$\cite{Potoff:1998}.  In our simulations, the solvent and
the nanoparticle comprise 107233 and 767 atoms, respectively,
corresponding to a length $L\approx 51$ of the periodic simulation
cell and a particle radius $R\approx 5$ (see \fref{fig:screenshot})
for a screenshot).  Concerning finite-size effects, which are mainly
due to the long range hydrodynamic interactions between the periodic
image particles, we refer the reader to the \SUPMAT, where we also
give some details concerning the home-grown code and its parallel
processing on graphics cards (GPUs).

In a typical simulation, the system was first equilibrated in the NPT
ensemble at the prescribed temperature of $T=0.75$ and pressure of
$P=0.01$ using a Nos\'{e}--Hoover thermostat and barostat. During the
subsequent heating of the nanoparticle to the temperature
$T_\mathrm{p}$, realised by rescaling the velocities of the atoms
belonging to the nanoparticle in each time step, the system evolved in
the NPH ensemble, while the particles near the boundary of the
simulation box were maintained at $T=0.75$.  At least four independent
trajectories of $2\cdot 10^7$ time steps $\delta t=0.005$
(corresponding to a physical duration of $100$ ns) were computed for
each nanoparticle temperature $T_\mathrm{p}$. Coordinates and momenta
were recorded after attaining stationary thermal conditions.  Note
that different physical heating mechanisms for the colloidal particle,
such as a heat source residing within the particle (which does not
directly affect its center-of-mass velocity) versus an external heat
supply (which generally does), correspond to different numerical
heating procedures.

\begin{figure}[t]
  \centering
  \includegraphics[width=\linewidth]{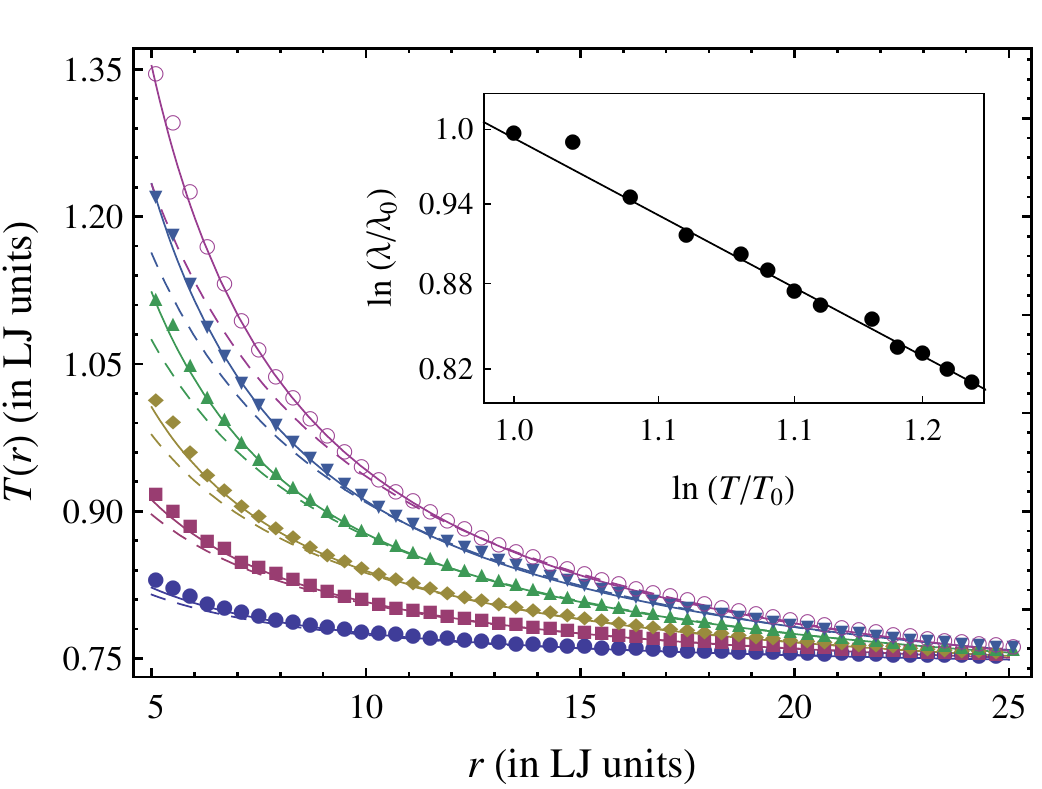}  
  \caption[For LoF]{Radial temperature profiles in the cool solvent
    around a hot nanoparticle of radius $R\approx 5$ for the particle
    temperatures $T_\mathrm{p}=1.00$~{\Large\color{myblue}$\bullet$},
    $1.25$~{\color{mypurple}{\small $\blacksquare$}},
    $1.50$~{\color{myokker}{\Large$\diamond$}},
    $1.75$~{\color{mygreen}$\blacktriangle$},
    $2.00$~{\color{mymarine}$\blacktriangledown$},
    $2.25$~{\color{mypurple}{\Large $\circ$}}. Dashed and solid lines
    are fits obtained from Fourier's law for a thermal conductivity
    $\lambda\,$=constant (dashed), as assumed in theory, and
    $\lambda(T)\propto 1/T$, as inferred from the fit to independent
    simulations of an isothermal bulk fluid (inset), respectively.}
  \label{fig:fig1}
\end{figure}

\begin{figure}[t]
  \includegraphics[width=\linewidth]{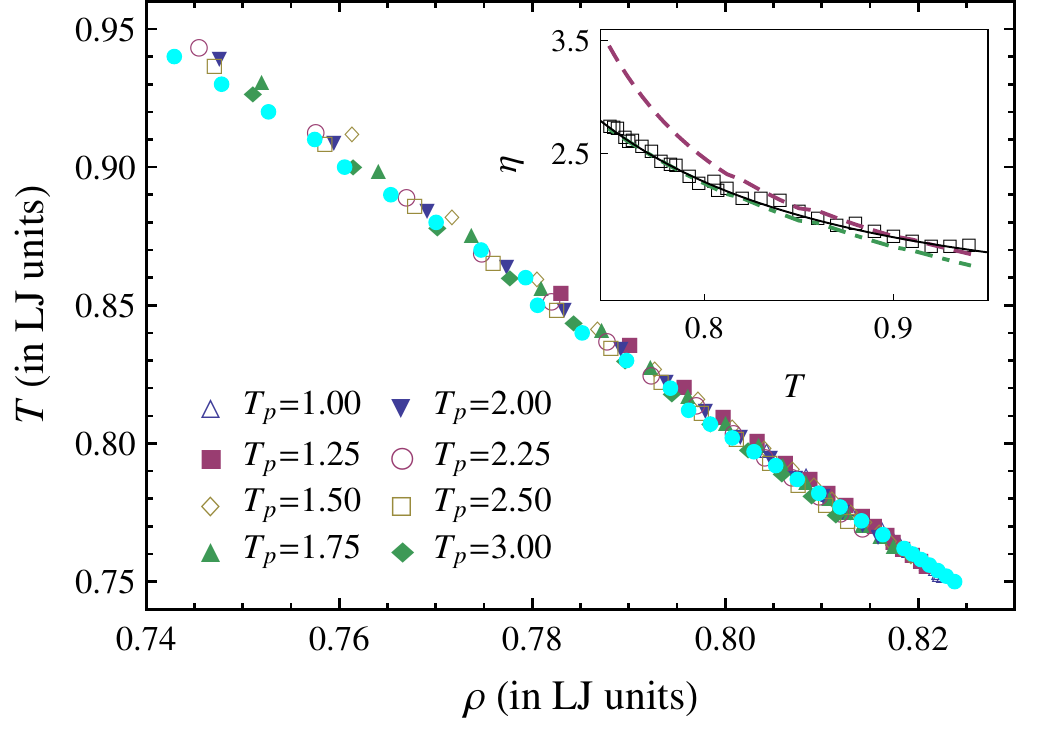}
  \caption{Sample points of the radial temperature and density
    profiles $T(r)$ and $\rho(r)$ around a hot nanoparticle maintained
    at various temperatures $T_\mathrm{p}$, in the $T-\rho$ plane. The
    filled circles ({\Large\color{mycyan}$\bullet$}) correspond to
    states for which the viscosity $\eta(T)$ of the bulk
    Lennard-Jones fluid was determined in separate simulations. The
    inset compares the latter($\square$) to the empirical formula
    from \fref{eq:etaT_fit} ({\bf ---}) and alternative expressions
    proposed in the recent literature \cite{Rowley:1997}
    ({\color{mypurple}$\mathbf{--}$}) \cite{Galli'ero:2005}
    ({\color{mygreen}$\mathbf{-\cdot-}$}).}
  \label{fig:fig2}
\end{figure}

Due to the finite compressibility of the Lennard-Jones fluid, the
simulation does not comply with the idealizations made in our
theoretical calculations (incompressible solvent, constant heat
conductivity). Therefore, to accurately describe the temperature
profile in the Lennard-Jones solvent (excluding the discontinuity due
to the Kaptiza resistance at the particle surface
\cite{Merabia:2009}), we need to account for the temperature
dependence of the thermal conductivity. Separate simulations of an
isothermal bulk fluid along the state-space curve depicted in
\fref{fig:fig2} (discussed below) yield an inverse temperature
dependence for the thermal conductivity, $\lambda\propto 1/T$, as
demonstrated in the inset of \fref{fig:fig1}. The main figure compares
the corresponding solution of the heat equation,
\begin{equation}
  \label{eq:Tr}
  T(r)=T_0 \left(1+\Delta T/T_0\right)^{\frac{R}{r}}\;,
\end{equation}
to the measured temperature profiles around the hot nanoparticle.
Here, $T_0\approx 0.75$ denotes the nominal ambient temperature at
infinite distance $r\to \infty$ from the particle.

\revision{While $T(r)$ can be obtained by averaging the kinetic
energy of the solvent particles in the vicinity of $\vec{r}$, it is
more subtle to deduce the local viscosity $\eta(r)$ from the available
microscopic data.} Under isothermal conditions, the viscosity can be
computed from the microscopic stress tensor $\sigma_{xy}$ via the
Green--Kubo formula $\eta = V(k_\mathrm{B} T)^{-1}\int_0^\infty
\langle \sigma_{xy}(t) \sigma_{xy}(0) \rangle\,\upd t$.  However, the
strong temperature gradient in our system precludes the evaluation of
the correlation function over a sufficiently large volume to get
reliable results.  We therefore evaluated the isothermal viscosity
$\eta(T)$ of a homogeneous bulk Lennard-Jones fluid with the intention
to translate it into $\eta(r)=\eta[T(r)]$ using $T(r)$ from
\fref{eq:Tr}. We observed that the radial dependence of the density
$\rho$ around the heated nanoparticle, which is neglected in the
analytical calculations, can be quite substantial for the
Lennard-Jones fluid. Plotting sample points of the radial temperature
and density profiles $T(r)$ and $\rho(r)$ around a heated nanoparticle
in the $T-\rho$ plane for various particle temperatures produced the
state curve delineated by the data points in \fref{fig:fig2}. In
determining the isothermal bulk viscosity $\eta(T)$ of the
Lennard-Jones fluid using to the Green--Kubo formula, we therefore
took care to vary the barostat pressure such as to confine the
measured bulk states to this curve.  In the inset of \fref{fig:fig2},
we compare our data for $\eta(T)$ to a phenomenological two-parameter
fit
\begin{equation}
  \label{eq:etaT_fit}
  \ln\left[\eta(T)/\eta_\infty\right]=(A/T)^4 \;.
\end{equation} 

\begin{figure}[tb]
  \includegraphics[width=\linewidth]{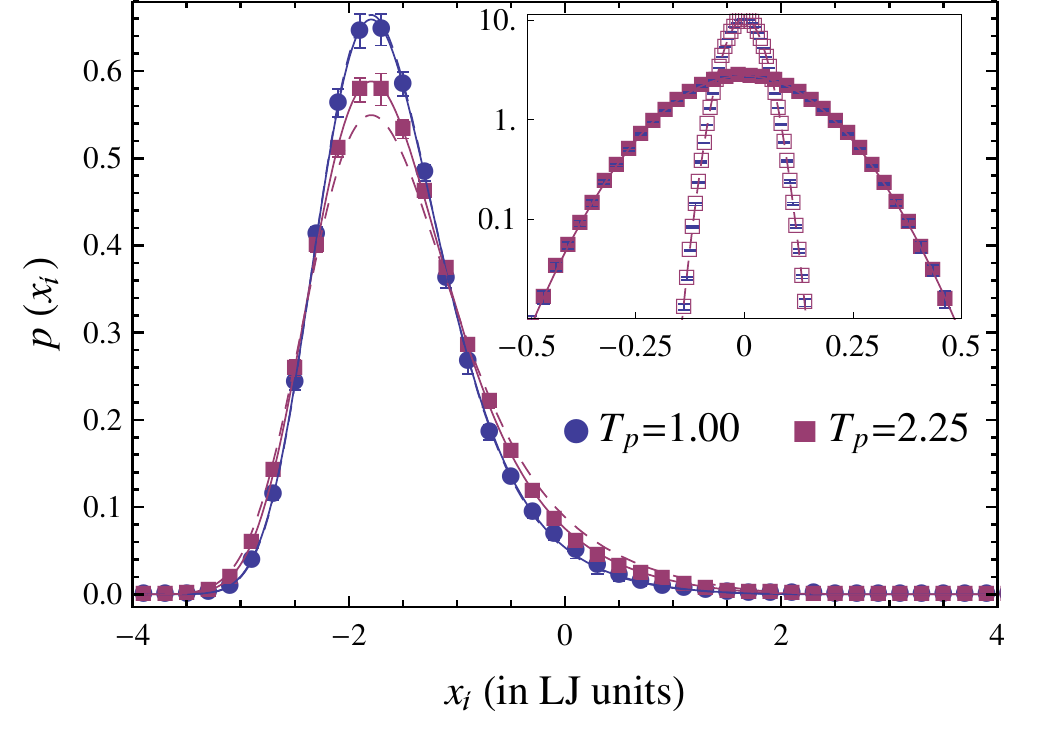} 
  \caption{Distribution of the position coordinates of a hot Brownian
    particle in an asymmetric, anharmonic potential well for two
    nanoparticle temperatures. The Boltzmann distribution fits the
    data for the effective temperature $T_\mathrm{HBM}$ (solid line)
    but not for $T_{\mathrm k}$ (dashed line), as it should.  The
    inset depicts the distributions of the position (filled symbols)
    and velocity (open symbols) coordinates in a harmonic well of
    strength $K=50$ and Boltzmann distributions for the effective
    temperatures $T_\mathrm{HBM}$ (solid line) and $T_{\mathrm k}$
    (dashed line).}
  \label{fig:fig6}
\end{figure}

Other observables that can directly be obtained from the simulation
are the effective steady-state friction $\zetaHBM$ and diffusivity
$\DHBM$ of the hot Brownian particle. While good estimates for $\DHBM$
are deduced from the nanoparticle trajectories, determining the
friction is slightly more subtle \cite{Espanol:1993,Lee:2004}. We
inferred $\zetaHBM$ from the decay of the momentum auto-correlation
function of the nanoparticle using the Brownian limit
\cite{Bocquet:1994}. In the simulations, the measured force
$\mathbf{F}$ on the colloid does not correspond to the random force
$\boldsymbol{\xi}$ entering the Green--Kubo formula for the friction
coefficient.  From a generalised Langevin description, it can be shown
that the correlations $\langle \mathbf{F}(0)\mathbf{F}(t) \rangle $
and $\langle \boldsymbol{\xi}(0) \boldsymbol{\xi}(t)\rangle$ become
equal only in the limit of a diverging reduced mass $\mu \to
\infty$. The Brownian limit amounts to first taking the mass of the
colloidal particle to infinity and then taking the thermodynamic limit
of infinitely many solvent particles. As a consequence, the momentum
relaxation time of the colloid becomes large compared to the typical
relaxation time of the random force autocorrelations, and its momentum
autocorrelation function becomes Markovian \cite{Espanol:1993}. It is then found to
exhibit an exponential decay
\begin{equation}
\label{p_decay}
\langle \vec P(t) \vec P(0)\rangle =\langle \vec P^2(0)\rangle
e^{-(\zetaHBM/\mu)t}
\end{equation}
with the decay time determined by the friction coefficient $\zetaHBM$
and the reduced mass $\mu$. In practice, the Brownian limit is
realised following the constrained-dynamics approach
\cite{Hijon:2010}.

\begin{figure}[t]
  \includegraphics[width=\linewidth]{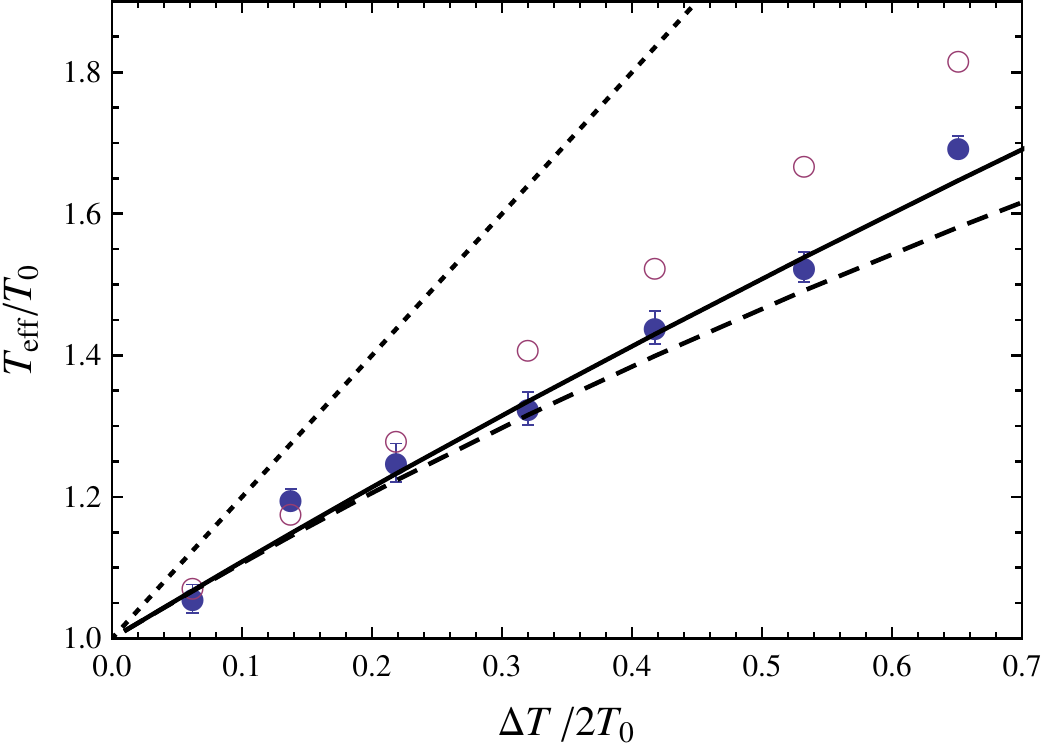}
  \caption{Effective temperatures of translational hot Brownian
    motion. \emph{Simulation:} $\THBM$ ({\Large
      \color{myblue}$\bullet$}) from the generalised Einstein
    relation, \fref{eq:GER}; apparent equipartition temperature
    $T_\mathrm{k}$ ({\color{mypurple}{\Large $\circ$}}) for the
    particle velocity; solvent temperature at the particle surface
    (dotted) \emph{Theory:} $\THBM$ according to
    \fref{eq:T_HBM_Manuel} (solid line) and the previous thermodynamic
    estimate, eq.~(22) of Ref.~\cite{Rings:2011} (dashed line), both
    evaluated (assuming an incompressible fluid) with the exact
    numerical differential shell method \cite{Rings:2011}.}
  \label{fig:fig3}
\end{figure}

Knowing $\zetaHBM$ and $\DHBM$, we can determine the configurational
effective temperature $\THBM$ using the generalised Einstein relation
\fref{eq:GER}. The temperature $T_\mathrm{k}$ for the kinetic degrees
of freedom is extracted from the equal-time velocity autocorrelation
of the nanoparticle by applying the equipartition relation $\mu\langle
\vec U^2 \rangle= 3k_BT_\mathrm{k}$. Note, however, that for hot
particles, which are far from equilibrium, the pertinence of the
equipartition temperature $T_\mathrm{k}$ is \emph{a priori}
unclear. We refer the reader to Joly \etal~\cite{Joly:2011} for a
discussion of this aspect.

\textbf{Results and discussion} In the remainder of this paper, we
present our simulation results for the effective temperatures $\THBM$
and $T_\mathrm{k}$ of translational hot Brownian motion and test our
theoretical predictions.

To support the first claim formulated in the introduction, we
performed simulations in the presence of external potential forces
derived from the harmonic potential $\mathcal{V}(\vec r)=K r^2/2$ for
different values of the stiffness parameter $K$. Generalising the
corresponding definitions for the free particle, the effective
temperatures are inferred from the particle's mean kinetic energy
($k_\mathrm{B}T_\mathrm{k}\equiv \mu \langle \mathbf{U}^2\rangle /3$)
and mean square displacement ($k_\mathrm{B}\THBM \equiv K \langle
\mathbf{r}^2 \rangle/3$) in the harmonic well. In the \SUPMAT, they
are compared to results for a free Brownian particle maintained at the
same temperature to demonstrate perfect agreement to within the
statistical errors. In the inset of \fref{fig:fig6} it is moreover
verified that the coordinates and momenta are indeed Boltzmann
distributed with the respective effective temperatures. The main plot
demonstrates for two exemplary particle temperatures $T_\mathrm{p}$
that this generalises to an asymmetric and anharmonic potential well
$\mathcal{V}(\vec r)=\sum_{i=1}^3v(x_i)$ with $v(x) =K x^4/4 + b
x$. Altogether, the good agreement in all cases confirms that the
particle velocity and position are distributed according to effective
Boltzmann distributions governed by the same effective temperatures
$T_\mathrm{k}$ and $\THBM$ as found for the freely diffusing particle.

To establish also our second claim, \fref{fig:fig3} provides a
comparison of our theoretical prediction for $\THBM$ from
\fref{eq:T_HBM_Manuel} and the simulation results obtained by means of
the generalised Einstein relation, \fref{eq:GER}. The good agreement
validates \fref{eq:GER} and \fref{eq:T_HBM_Manuel} over a wide
temperature range. \revision{Also note that, at strong heating,
  \fref{eq:T_HBM_Manuel} fits the data significantly better than the
  previous thermodynamic estimate, eq.~(22) of
  Ref.~\cite{Rings:2011}---which amounts to a weighted average of $1/T$
  instead of $T$ in \fref{eq:T_HBM_Manuel}, and is therefore
  systematically too small.}  As detailed in the \SUPMAT, the
configurational temperature $\THBM$ is found to be insensitive to the
precise microscopic conditions such as the particle solubility or the
physical realisation of the heating mechanism (internal/external),
while $T_\mathrm{k}$ is sensitive to the heating mechanism.

In summary, more than a century after Langevin published his famous equation
\cite{Langevin:1908}, we can now describe the overdamped Brownian
motion of a hot nanoparticle by
\begin{equation}
  \zetaHBM \dot{\vec r} = -\nabla \mathcal{V}(\vec r) +
  \boldsymbol{\xi}(t) \;.
\end{equation}
The effective Gaussian thermal noise $\xi(t)$ is characterised by
$\langle\xi(t)\rangle=0$ and
$\langle\xi(t)\xi(t')\rangle=2\THBM\zetaHBM\delta(t-t')$. All
coefficients are explicitly known over a wide temperature range, and
accurate analytical expressions for them have been derived. Most
importantly, we have established the following fundamental properties
of the effective temperature $\THBM$ given in \fref{eq:T_HBM_Manuel}:
(1) it governs the Boltzmann factor for the probability distribution
of a hot nanoparticle in an external potential $\mathcal{V}(\vec r)$;
(2) it may be expressed in terms of the directly measurable effective
diffusivity $\DHBM$ and friction $\zetaHBM$ via the generalised
Einstein relation, \fref{eq:GER}; and (3) it takes distinct values for
translational and rotational degrees of freedom
\cite{Rings:Rotational:2011}. The momenta of the particle were also
found to be Boltzmann distributed with and without external potential
forces, but according to yet another effective temperature
$T_\mathrm{k}$ that also takes distinct values for translational and
rotational degrees of freedom, and, in contrast to $\THBM$, is
sensitive to the precise heating mechanism. Altogether, our findings
strongly support the expectation \cite{Rings:2010} that the effective
temperature $\THBM$ can be treated as a \emph{bona-fide} temperature
for the configurational Brownian motion of an individual hot particle,
and thus pave the way for manifold applications.  It remains as an
intriguing open question how far this convenient description can be
extended to account for finite particle densities, anisotropies,
\cite{Ruijgrok:2011} and self-thermophoresis \cite{Golestanian:2010}.

\textbf{Supporting Information}
A PDF file with technical and supplementary information containing a
detailed formulation of the contraction of the fluctuating
hydrodynamics problem to a Markovian Langevin description of the
Brownian motion of the nanoparticle, convenient analytical
approximations for $\THBM$ and $\DHBM$, a discussion of how to obtain
$\zetaHBM$ and how to deal with finite size effects in the numerical
simulations, information about the parallel processing of the
simulation code, and various supporting plots is available free of
charge via the Internet at http://arxiv.org/...

\acknowledgments
We gratefully acknowledge helpful discussions with Jean-Louis Barrat
(Grenoble), Ramin Golestanian (Oxford), and Markus Selmke (Leipzig),
and thank Hugo Brandao for a careful reading of the manuscript. This
work was supported by the Alexander von Humboldt foundation, the
Deutsche Forschungsgemeinschaft (DFG) via FOR 877 and, within the
German excellence initiative, the Leipzig School of Natural Sciences
``Building with molecules and nano-objects''.

\bibliography{library}

\begin{thebibliography}{10}
\expandafter\ifx\csname url\endcsname\relax\def\url#1{\texttt{#1}}\fi

\bibitem{Rings:2010}
\Name{Rings D., Schachoff R., Selmke M., Cichos F. \and Kroy K.} \REVIEW{Phys.
  Rev. Lett. }{105}{2010}{090604}.

\bibitem{Ruijgrok:2011}
\Name{Ruijgrok P.~V., Verhart N.~R., Zijlstra P., Tchebotareva A.~L. \and Orrit
  M.} \REVIEW{Phys. Rev. Lett. }{}{2011}{} accepted Wednesday May 18, 2011.

\bibitem{Jiang:2010}
\Name{Jiang H.-R., Yoshinaga N. \and Sano M.} \REVIEW{Phys. Rev. Lett.
  }{105}{2010}{268302}.

\bibitem{Lasne:2006}
\Name{Lasne D., Blab G.~A., Berciaud S., Heine M., Groc L., Choquet D., Cognet
  L. \and Lounis B.} \REVIEW{Biophys. J. }{91}{2006}{4598}.
\newline\url{http://dx.doi.org/10.1529/biophysj.106.089771}

\bibitem{vanDijk:2006}
\Name{van Dijk M.~A., Tchebotareva A.~L., Orrit M., Lippitz M., Berciaud S.,
  Lasne D., Cognet L. \and Lounis B.} \REVIEW{Phys. Chem. Chem. Phys.
  }{8}{2006}{3486}.
\newline\url{http://dx.doi.org/10.1039/b606090k}

\bibitem{octeau-etal:2009}
\Name{Octeau V., Cognet L., Duchesne L., Lasne D., Schaeffer N., Fernig D.~G.
  \and Lounis B.} \REVIEW{ACS Nano }{3}{2009}{345}.
\newline\url{http://dx.doi.org/10.1021/nn800771m}

\bibitem{Huang:2010}
\Name{Huang X. \and El-Sayed M.~A.} \REVIEW{Journal of Advanced Research
  }{1}{2010}{13 }.
\newline\url{http://www.sciencedirect.com/science/article/B9HCY-4YCHFV0-5/2/0a%
1369e79150fec1ca6b79a2ca235b10}

\bibitem{Gaiduk:2010}
\Name{Gaiduk A., Yorulmaz M., Ruijgrok P.~V. \and Orrit M.} \REVIEW{Science
  }{330}{2010}{353}.
\newline\url{http://www.sciencemag.org/cgi/content/abstract/330/6002/353}

\bibitem{bericiaud-etal:2004}
\Name{Berciaud S., Cognet L., Blab G.~A. \and Lounis B.} \REVIEW{Phys. Rev.
  Lett. }{93}{2004}{257402}.
\newline\url{http://link.aps.org/abstract/PRL/v93/e257402}

\bibitem{radunz-etal:2009}
\Name{Rad{\"u}nz R., Rings D., Kroy K. \and Cichos F.} \REVIEW{J.\ Phys.\
  Chem.\ A }{113}{2009}{1674}.
\newline\url{http://pubs.acs.org/doi/abs/10.1021/jp810466y}

\bibitem{Urban:2009}
\Name{Urban A.~S., Fedoruk M., Horton M.~R., Rädler J.~O., Stefani F.~D. \and
  Feldmann J.} \REVIEW{Nano Lett. }{9}{2009}{2903} pMID: 19719109.
\newline\url{http://pubs.acs.org/doi/abs/10.1021/nl901201h}

\bibitem{Qian:2008}
\Name{Qian X., Peng X.-H., Ansari D.~O., Yin-Goen Q., Chen G.~Z., Shin D.~M.,
  Yang L., Young A.~N., Wang M.~D. \and Nie S.} \REVIEW{Nat Biotech
  }{26}{2008}{83}.
\newline\url{http://dx.doi.org/10.1038/nbt1377}

\bibitem{Kyrsting:2011}
\Name{Kyrsting A., Bendix P.~M., Stamou D.~G. \and Oddershede L.~B.}
  \REVIEW{Nano Letters }{11}{2011}{888}.
\newline\url{http://pubs.acs.org/doi/abs/10.1021/nl104280c}

\bibitem{Rings:2011}
\Name{Rings D., Selmke M., Cichos F. \and Kroy K.} \REVIEW{Soft Matter
  }{7}{2011}{3441}.
\newline\url{http://dx.doi.org/10.1039/C0SM00854K}

\bibitem{Rings:Rotational:2011}
\Name{Rings D., Chakraborty D., Cichos F. \and Kroy K.} \Book{Rotational {H}ot
  {B}rownian {M}otion} (to be published).

\bibitem{Joly:2011}
\Name{Joly L., Merabia S. \and Barrat J.-L.} \REVIEW{EPL }{94}{2011}{50007}.
\newline\url{http://dx.doi.org/10.1209/0295-5075/94/50007}

\bibitem{haw_middle-world:2006}
\Name{Haw M.} \Book{Middle World: The Restless Heart of Matter and Life}
  (Macmillan, New York) 2006.

\bibitem{Seifert:2008}
\Name{Seifert U.} \REVIEW{Eur.\ Phys.\ J.\ B }{64}{2008}{423}.
\newline\url{http://dx.doi.org/10.1140/epjb/e2008-00001-9}

\bibitem{Fox:1970a}
\Name{Fox R.~F. \and Uhlenbeck G.~E.} \REVIEW{Phys. Fluids }{13}{1970}{1893}.
\newline\url{http://link.aip.org/link/?PFL/13/1893/1}

\bibitem{Fox:1970b}
\Name{Fox R.~F. \and Uhlenbeck G.~E.} \REVIEW{Phys. Fluids }{13}{1970}{2881}.
\newline\url{http://link.aip.org/link/?PFL/13/2881/1}

\bibitem{hauge-martin_lof:73}
\Name{Hauge E.~H. \and Martin-L\"of A.} \REVIEW{J.\ Stat.\ Phys.
  }{7}{1973}{259}.
\newline\url{http://dx.doi.org/10.1007/BF01030307}

\bibitem{Bocquet:1994}
\Name{Bocquet L., Hansen J.-P. \and Piasecki J.} \REVIEW{J. Stat. Phys.
  }{76}{1994}{527} 10.1007/BF02188674.
\newline\url{http://dx.doi.org/10.1007/BF02188674}

\bibitem{Keblinski:2006}
\Name{Keblinski P. \and Thomin J.} \REVIEW{Phys. Rev. E }{73}{2006}{010502}.
\newline\url{http://link.aps.org/doi/10.1103/PhysRevE.73.010502}

\bibitem{Li:2009}
\Name{Li Z.} \REVIEW{Phys. Rev. E }{80}{2009}{061204}.

\bibitem{Shin:2010}
\Name{Shin H.~K., Kim C., Talkner P. \and Lee E.~K.} \REVIEW{Chem. Phys.
  }{375}{2010}{316} stochastic processes in Physics and Chemistry (in honor of
  Peter Hänggi).
\newline\url{http://www.sciencedirect.com/science/article/B6TFM-5046MF6-6/2/35%
890b8275d1acc8d65c0d5fe41ed815}

\bibitem{supmat}
{S}upplementary materials are available online.
\newline\url{http://arxiv.org...}

\bibitem{Frenkel:2002}
\Name{Frenkel D. \and Smit B.} \Book{Understanding Molecular Simulation: From
  Algorithms to Applications} 2nd Edition (Academic Press, Inc., San Diego,
  California, USA) 2002.

\bibitem{Potoff:1998}
\Name{Potoff J.~J. \and Panagiotopoulos A.~Z.} \REVIEW{The Journal of Chemical
  Physics }{109}{1998}{10914}.
\newline\url{http://link.aip.org/link/?JCP/109/10914/1}

\bibitem{Merabia:2009}
\Name{Merabia S., Keblinski P., Joly L., Lewis L.~J. \and Barrat J.-L.}
  \REVIEW{Phys. Rev. E }{79}{2009}{021404}.
\newline\url{http://pre.aps.org/abstract/PRE/v79/i2/e021404}

\bibitem{Rowley:1997}
\Name{Rowley R. \and Painter M.} \REVIEW{Int. J. Thermophys. }{18}{1997}{1109}
  10.1007/BF02575252.
\newline\url{http://dx.doi.org/10.1007/BF02575252}

\bibitem{Galli'ero:2005}
\Name{Galli\'{e}ro G., Boned C. \and Baylaucq A.} \REVIEW{Industrial \&
  Engineering Chemistry Research }{44}{2005}{6963}.
\newline\url{http://pubs.acs.org/doi/abs/10.1021/ie050154t}

\bibitem{Espanol:1993}
\Name{Espanol P. \and Zuniga I.} \REVIEW{The Journal of Chemical Physics
  }{98}{1993}{574}.
\newline\url{http://link.aip.org/link/?JCP/98/574/1}

\bibitem{Lee:2004}
\Name{Lee S.~H. \and Kapral R.} \REVIEW{The Journal of chemical physics
  }{121}{2004}{11163}.
\newline\url{http://www.ncbi.nlm.nih.gov/pubmed/15634070}

\bibitem{Hijon:2010}
\Name{Hijon C., Espanol P., Vanden-Eijnden E. \and Delgado-Buscalioni R.}
  \REVIEW{Faraday Discuss. }{144}{2010}{301}.
\newline\url{http://dx.doi.org/10.1039/B902479B}

\bibitem{Langevin:1908}
\Name{Langevin P.} \REVIEW{C. R. Acad. Sci. (Paris) }{146}{1908}{530}.
\newline\url{http://www.physik.uni-augsburg.de/theo1/hanggi/History/Langevin19%
08.pdf}

\bibitem{Golestanian:2010}
\Name{Golestanian R.} \REVIEW{Physics }{3}{2010}{108}.

\end{thebibliography}
\bibliographystyle{eplbib}

\end{document}